\documentclass[useAMS,usenatbib]{pasj02}

\usepackage{tikz}
\usepackage{lscape}
\usepackage{color}
\usepackage{comment}
\usepackage{graphicx}
\usepackage{natbib, aas_macros}
\usepackage{bm}
\usepackage{placeins}
\usepackage{amsmath}
\usepackage{cuted}
\usepackage{flushend}

\usepackage{soul}
\sethlcolor{pink}
\usepackage[switch,mathlines]{lineno}

\newcommand{\simgt}{\lower.5ex\hbox{$\; \buildrel > \over \sim \;$}}
\newcommand{\simlt}{\lower.5ex\hbox{$\; \buildrel < \over \sim \;$}}

\def\h70kpc{\mathrel{h_{70}^{-1}{\rm kpc}}}

\def\h70Msol{\mathrel{h_{70}^{-1}M_\odot}}

\usepackage{version}

\begin{document}

\Received{}
\Accepted{}



\Received{}
\Accepted{}

\Received{}
\Accepted{}

\title{Discovery of an X-ray bridge between the comma-shaped gas and the main cluster in MCXC J0157.4-0550\thanks{Based on observations obtained with XMM-Newton, an ESA science mission with instruments and contributions directly funded by ESA Member States and NASA. Based on data collected at Subaru Telescope, which is operated by the National Astronomical Observatory of Japan.}}


\author{Chong \textsc{Yang}\altaffilmark{1}}
\email{you@astro.hiroshima-u.ac.jp}
\altaffiltext{1}{Physics Program, Graduate School of Advanced Science and Engineering, Hiroshima University, 1-3-1 Kagamiyama, Higashi-Hiroshima, Hiroshima 739-8526, Japan}

\author{Nobuhiro \textsc{Okabe}\altaffilmark{1,2,3}}
\email{okabe@hiroshima-u.ac.jp}
\altaffiltext{2}{Hiroshima Astrophysical Science Center, Hiroshima University, 1-3-1 Kagamiyama, Higashi-Hiroshima, Hiroshima 739-8526, Japan}
\altaffiltext{3}{Core Research for Energetic Universe, Hiroshima University, 1-3-1, Kagamiyama, Higashi-Hiroshima, Hiroshima 739-8526, Japan}

\author{Yasushi \textsc{Fukazawa}\altaffilmark{1,2,3}}


\maketitle

\begin{abstract}
We report the discovery of a faint X-ray bridge connecting between the comma-shaped gas and the main cluster in MCXC J0157.4-0550, using {\it XMM-Newton} image. The filamentary structure is found in a model-independent manner in both topological features and Gaussian Gradient Magnitude filtering. The X-ray surface brightness profile perpendicular to the filament is detected at a $5.5\sigma$ level. Weak-lensing (WL) analysis using the Subaru/HSC-SSP Survey archive data strongly supports the two mass components. Given a prior from the stellar masses, we obtain $M_{200}^{\rm main}=2.68_{-0.92}^{+1.11}\times 10^{14}\,h_{70}^{-1}M_\odot$ and $M_{200}^{\rm sub}=0.46_{-0.22}^{+0.38}\times 10^{14}\,h_{70}^{-1}M_\odot$. The main axis of the projected halo distribution is more likely to align with the direction of the main cluster than to be oriented perpendicularly. Similar X-ray distributions have been identified in the literature on numerical simulations. The filamentary structure forms in the following manner: as the gas is stripped by ram pressure near the pericenter, it gets dragged by tidal rotation. Once free from this rotation, the gas moves inertially in a direction parallel to the tangential velocity at the pericenter. The comma-shaped gas, with tails pointing in the opposite direction to the main cluster, is also formed by the current tidal rotation as it moves away from the main cluster.
This warrants us that, although it is sometimes thought based on the X-ray morphology alone that the tail is pointing in the opposite direction to the merger motion, this is not necessarily the case. The information of the X-ray filamentary remnant from the cluster merger, together with the 2D WL shear data, provides constraints on the merger parameters, indicating an infalling velocity of approximately $1000\, {\rm km\, s^{-1}}$ and an impact parameter of $0.9$ Mpc.
\end{abstract}

\clearpage %


\section{Introduction}

Galaxy clusters are the unique laboratory for understanding the formation of hierarchical structures, because galaxy clusters are the most massive objects in the Universe, and then less massive objects sometimes fall into main halos. Such clusters are known as merging clusters. Since the energy release of the gravitational potential is huge, various non-linear phenomena in the intracluster medium are triggered due to its collisional nature \citep{2001ApJ...561..621R}.
X-ray observations of merging clusters have revealed these active features \citep{2007PhR...443....1M}, such as bow shocks \citep[e.g.][]{2002ApJ...567L..27M}, cold fronts \citep[e.g.][]{2000ApJ...541..542M}, and ram-pressure stripping \citep[e.g.][]{2016PASJ...68...85S}. 
These X-ray morphologies help us to infer the merging direction on the sky plane. Specifically, if the tail is pulled like a broom star, the direction of the greater tail spread is inferred to be behind the direction of motion on the sky plane.

As a similar feature of the tail, \citet{2018PASJ...70S..22M} discovered a comma-shaped gas distribution in MCXC J0157.4-0550 at $z=0.1289$. They inferred from its morphology the scenario that the gas is infalling with a large angular momentum.
This paper reports the detection of an X-ray bridge between the comma-shaped gas and the diffuse X-ray emission of the main cluster, compares it with numerical simulations, and discusses the merger geometry combined with weak-lensing analysis. 

This paper is organised as follows: Secs. \ref{sec:data} and \ref{sec:result} briefly describe the data analysis and present the results, respectively.  Secs \ref{sec:dis} and \ref{sec:summary} are devoted to discussion and summary, respectively.
We assume a flat $\Lambda$CDM cosmology with the cosmological parameters of $H_0=70\,{\rm {\rm km\,s^{-1}}Mpc^{-1}}$, $\Omega_{m,0}=0.3$, and $\Omega_{\Lambda,0}=0.7$. Throughout this paper, the confidence level for errors is $1\sigma$.

\section{Data analysis} \label{sec:data}

XMM-Newton data were analyzed with the ESAS (Extended Source Analysis Software)
package \citep{2008A&A...478..615S}. 
The details of the data analysis are described by \citet{2018PASJ...70S..22M} and \citet{2023MNRAS.520.6001P}.  The data were filtered
for intervals of high background due to soft proton flares, defined to be periods when the rates were outside the $2\sigma$ range of a rate distribution. The net exposure times are 27.8, 27.2, and 16.1 ksec for mos1, mos2, and pn, respectively.
For image analysis, point sources are removed from three EPIC (MOS1,
MOS2, and pn) images with simultaneous maximum likelihood PSF fitting. We filled point sources in the count image with random values from a Poisson distribution with a mean in the surrounding pixels within 2 times their radius excluding them. The count rate image is calculated by subtracting the number of particle background counts, model soft proton counts and model solar wind charge exchange counts from the count image and then dividing by the exposure map. We combine MOS1, MOS2 and PN images.
For spectral analysis, we followed \citet{2008A&A...478..615S}, where all spectra extracted from regions of interest are simultaneously fitted with a common model, including particle and cosmic background components which are assumed to be uniform across the detector except for instrumental lines. The instrumental background spectrum is modeled using data obtained with the closed filter wheel, accessible in ESAS CALDB, and is subsequently subtracted from the observed spectrum. The additional particle backgrounds, comprising a soft proton-induced continuum and instrumental lines, are characterized by incorporating a power-law spectrum and narrow Gaussian lines with fixed central energies into the fitting model, respectively. The cosmic diffuse background is composed of the cosmic X-ray background (CXB), Galactic diffuse emission, and emission lines resulting from solar wind charge exchange (SWCX). The CXB component is modeled with a power-law spectrum with a fixed index of 1.46 according to \citet{2008A&A...478..615S}. The Galactic diffuse emission is modeled by combining absorbed and unabsorbed thermal plasma emission models. We used a thermal plasma emission model, APEC \citep{2001ApJ...556L..91S}, with the Galactic photoelectric absorption model, phabs \citep{1992ApJ...400..699B}. We fixed the cluster redshift and the hydrogen column density of $N_H=2.34\times10^{20}\,{\rm cm^{-2}}$. We used xspec (v12.15.0) with the pgstat statistic.

We retrieved the archival HSC $i$'-band image and measured galaxy shapes of optical galaxies by \citet[][the KSB+ method]{1995ApJ...449..460K}. The details of the shape measurements are described by \citet{2016MNRAS.461.3794O}. 
The photometric data were retrieved from the archival HSC-SSP data \citep{2022PASJ...74..247A}. 
The photometric redshift of each galaxy was estimated from an average of 50 neighbouring COSMOS galaxies \citep{2013A&A...556A..55I} in the $g',r',i'$ and $z'$ plane. The background galaxies were selected in the color-color plane, following \citet{Medezinski18}. The resulting number density of the background galaxies is $n_g\simeq11\,{\rm arcmin}^{-2}$.

\section{Results} \label{sec:result}

\begin{figure}[ht]
\includegraphics[width=\hsize]{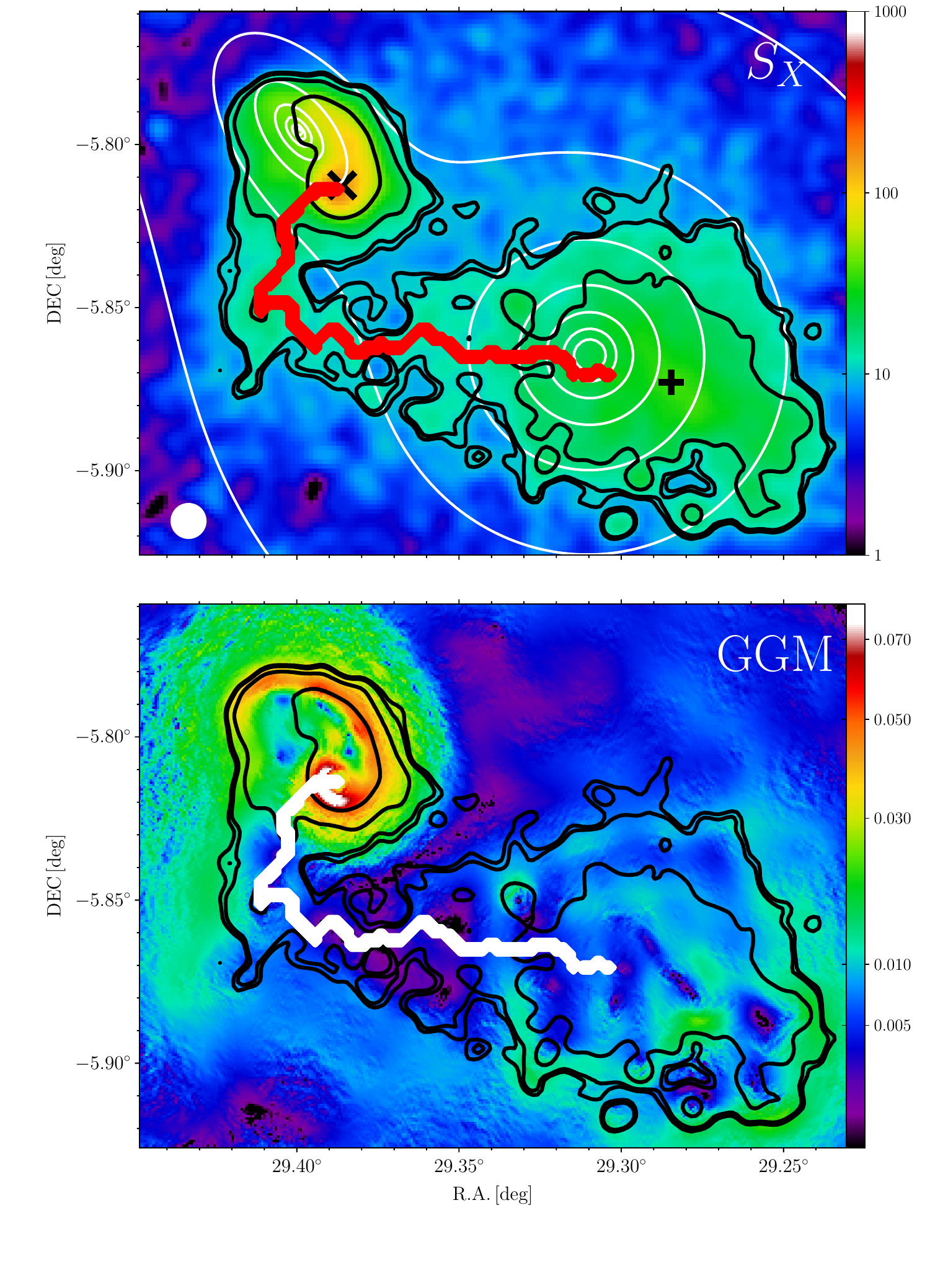}
\caption{{\it Top}: The {\it XMM-Newton} count rate image in 0.4-2.3 keV smoothed with a Gaussian kernel of FWHM$\simeq2\,[{\rm arcmin}]$. The black contours are represented by [10,16,34,64,106,160] in the unit of counts s$^{-1}$ deg$^{-2}$. The red line denotes the filament detected by DisPerSE. Black $+$ and $\times$ represent the brightest galaxies ($z=0.12823$ and $z=0.12869$) associated with each gas halo. The white contours are the modelled mass map with the best-fit parameters derived by 2D WL analysis with a prior of the mass ratio, spaced at square root intervals between [$0.02$,$1$] in $10^{15}\,h_{70}M_\odot{\rm Mpc}^{-2}$. The white circle in the left-bottom corner is the typical uncertainty of the centroid determinations. {\it Bottom}: Background color represents an adaptive GGM-filtered image. A weak edge is found at the south of the filament. {Alt text: Two stacked images showing the filament.}}
\label{fig:map}
\end{figure}

\begin{figure}[ht]
\includegraphics[width=\hsize]{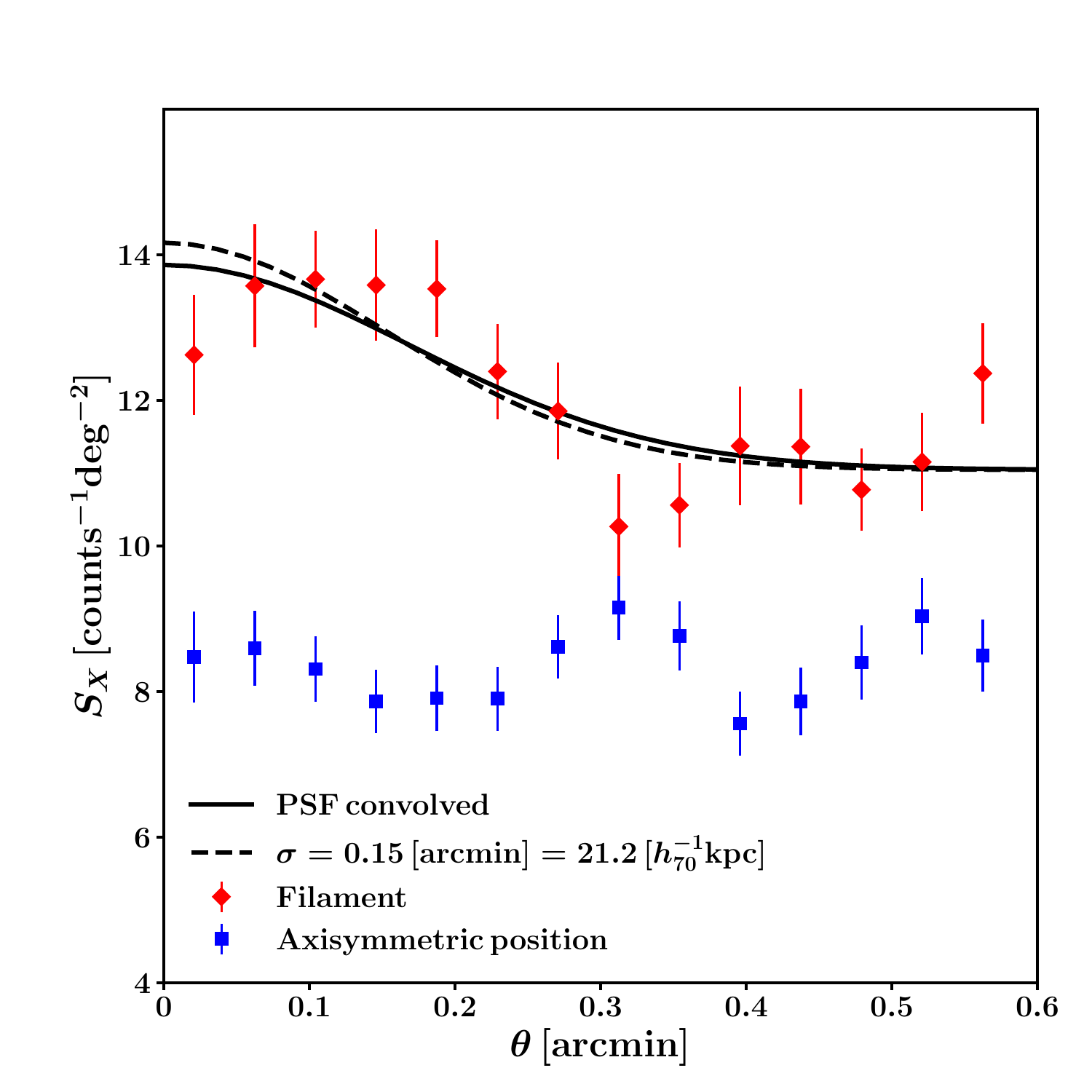}
\caption{The surface brightness profiles along a direction perpendicular to the filament (red diamonds) and to the axisymmetric position of the line connecting the two bright galaxies (blue squares). The PSF convolved and deconvolved Gaussian profiles are shown by the black solid and dashed lines, respectively. {Alt text: A graph showing the best-fitting results to the count rate across the filament, overplotted with the count rate at a non-filamentary position.}}
\label{fig:1dfilament}
\end{figure}

The top panel of Figure \ref{fig:map} shows the smoothed count rate image. The smoothing scale is FWHM$\simeq2\,{\rm arcmin}$.
We find a filamentary structure connecting the main and sub clusters. To quantify the filamentary structure, we run DisPerSE \citep{2011MNRAS.414..350S,2011MNRAS.414..384S} which identifies persistent topological features in a model-independent way. The detected filament is shown by the red line in the top panel of Figure \ref{fig:map}. The filament extends to the east from the main cluster, turning 90 degrees around $(\alpha, \delta)=(29.411^\circ,-5.848^\circ)$ and heading north to connect with the sub-cluster. We compute the surface brightness along a direction perpendicular to the filament. Here the filament chooses a region ($29.387^\circ \le \alpha \le 29.336^\circ$) where contamination of the main and sub clusters are minimal. The same calculation is performed at the axisymmetric position of the line connecting the two brightest galaxies (BCG; $+$ and $\times$ in the top panel of  Figure \ref{fig:map}) as a control. Figure \ref{fig:1dfilament} shows that the surface brightness has an excess at $\theta\simlt 0.3$ arcmin from the filament and becomes flat with increasing distance. 
In contrast, the surface brightness of the axisymmetric positions is lower and almost constant. We fit the surface brightness profile with a Gaussian profile ($\propto \exp\left[-\theta^2/(2\sigma^2)\right]$) added to the constant component, taking into account the PSF convolution. The resulting profiles with and without the PSF convolution are shown by the solid and dashed black lines, respectively. The scale parameter of the filament, $\sigma$, is $\sim21$ kpc. We compute a significance level of the filamentary structure by $S/N=\sqrt{\sum (S_{X}^{\rm fil}/S_X^{\rm con})^2}=5.5\pm1.7$, where $S_{X}^{\rm fil}$ and $S_X^{\rm con}$ are the surface brightness profiles of the filament and the control sample, respectively.

We make an adaptively smoothed image with a Gaussian gradient magnitude (GGM) filter \citep{2016MNRAS.457...82S,2016MNRAS.460.1898S,2022A&A...661A..36S}, incorporating the contour binning \citep{2006MNRAS.371..829S}, to detect an edge of the filamentary structure. The bottom panel of Figure \ref{fig:map} shows the GGM image in colour. The adaptive smoothing scales are determined with $S/N=60$. The weak edge bridging the main and sub clusters lies to the south of the filament, where the contamination from the main cluster X-ray emission is insignificant. The orientation angle from north to east is $\psi_{\rm fil}\simeq62\pm3$ deg, which is similar to the filament angle in the R.A. range computing the 1D filament profile (Figure \ref{fig:1dfilament}). When we change a parameter of $S/N$ in the GGM filter, the result does not change.

We measure weak-lensing masses for the main and sub clusters by two-dimensional (2D) analysis using the shape data within 2.5$\,h_{70}^{-1}$Mpc from the BCG \citep[e.g.][]{2011ApJ...741..116O,2021MNRAS.501.1701O,2025A&A...700A..46O}. We adopt an NFW model \citep{NFW96}, $\rho \propto 1/(r/r_s)/(1+r/r_s)^2$, where $r_s$ is a scale parameter. We assume the mass and concentration relation \citep{Bhattacharya13}, and treat $M_{\Delta}$ and their centers (${\alpha}_{\rm main},{\delta}_{\rm main}, {\alpha}_{\rm sub},{\delta }_{\rm sub}$) as free parameters, where $\Delta=200$ denotes that the mean mass density is $\Delta$ times of the critical density of the Universe. A log-likelihood of the 2D WL analysis  is described by ${\mathcal L}_{\rm 2DWL} \propto -0.5 \sum_{\alpha,\beta=1}^{2}\sum_{n,m}(\Delta \Sigma_{\alpha,n} - g_{{\rm NFW},\alpha}({\bm \theta}_n))C_{s,\alpha\beta,nm}^{-1} (\Delta
 \Sigma_{\alpha,m} - g_{{\rm NFW},\beta}({\bm \theta}_m))$, where $\Delta \Sigma_{\alpha}$ is the $\alpha=1,2$ component of a dimensional ellipticity at each spatial grid ($n,m$), $C_{s,\alpha\beta,nm}=\sigma_{\Delta \Sigma,mn}^2\delta_{\alpha\beta}$ is the shot noise variance of the dimensional ellipticity, and $g_{\rm NFW,\alpha}$ is a dimensional reduced shear of the NFW model at each position ${\bm \theta}$, respectively. 
We use a Markov Chain Monte Carlo (MCMC) method. The resulting WL masses are $M_{200}^{\rm main}=1.79_{-0.80}^{+1.61}\times 10^{14}\,h_{70}^{-1}M_\odot$  and $M_{200}^{\rm sub}=1.17_{-0.43}^{+0.92}\times 10^{14}\,h_{70}^{-1}M_\odot$ (Table \ref{tab:mass}). The mass ratio is $f_{\rm sub,200}=M_{200}^{\rm sub}/M_{200}^{\rm main}=0.64\pm0.47$. Due to the large uncertainty of the WL mass measurement, we could not conclude the major merger (the mass ratio is 1/2 $\sim$ 1/3) or minor merger (the mass ratio is less than 0.1). 

We estimate the mass by applying the multivariate scaling relations associated with stellar masses \citep{2022PASJ...74..175A}, because their quantities and distributions are not significantly changed by cluster mergers due to their non-collisional characteristics \citep{2019PASJ...71...79O}. 
The scaling relation can be described as ${\bm y}={\bm \alpha}+{\bm \beta}\ln M_{500}E(z)$, with ${\bm y}=(\ln M_s E(z),\ln M_{\rm BCG} E(z))$. Here, $M_s$ and $M_{\rm BCG}$ represent the stellar mass of member galaxies of the BCGs, respectively.  The stellar masses are calculated using the same procedure as explained in \citet{2022PASJ...74..175A}. The mass profiles of stellar and gas mass for each component, centered on the BCGs, are evaluated using a synthetic model of the central and off-center components. The stellar mass distribution is presumed to follow a power law. Then, we iteratively solve the following equation; $\ln f_{\rm sub,500}= ({\bm \beta}^T {\bm C}^{-1}{{\bm y}_{\rm ratio}}/({\bm \beta}^T {\bm C}^{-1}{\bm \beta})$ with ${\bm C}={\bm C}_{\rm int}+{\bm \sigma_\beta}^2(\ln f_{\rm sub,500})^2$ where ${\bm C}_{\rm int}$ is the intrinsic covariance, ${\bm \sigma}_{\bm \beta}$ is the measurement errors of the slope parameter, ${\bm y}_{\rm ratio}$ is the ratio of the bayronic components; ${\bm y}_{\rm ratio}=(\ln M_s^{\rm sub}(<r_{500}^{\rm sub})/M_s^{\rm main}(<r_{500}^{\rm main}), \ln  M^{\rm sub}_{\rm BCG}/M_{\rm BCG}^{\rm main})$, respectively. We assume the sum of $M_{200}$ to describe a total lensing signal and convert to $M_{500}=2.2\times 10^{14}\,h_{70}^{-1}M_\odot$ with the mass and concentration relation \citep{Bhattacharya13}. We derive an iterative solution for ${\bm y}_{\rm ratio}$, arriving at $(0.18\pm0.01,0.17)$, and establish $f_{\rm sub,500}=0.16\pm0.11$, subsequently estimating $f_{\rm sub,200}=0.15\pm0.11$. When we change by $\pm 1\sigma$ error of the total mass, a change of the mass ratio, $\delta f_{\rm sub,500}=^{-0.02}_{+0.01}$, is negligible. Given a Gaussian prior of the mass ratio $f_{\rm sub,200}=0.15\pm0.11$, we fit with the projected elliptical NFW model \citep{Oguri10b} with the projected halo ellipticity, $\varepsilon=1-b/a$, where $a$ and $b$ are the major and minor axes of the mass distribution on the sky plane, respectively. The resulting WL masses (Table \ref{tab:mass}) are $M_{200}^{\rm main}=2.68_{-0.93}^{+1.11}\times 10^{14}\,h_{70}^{-1}M_\odot$, $M_{200}^{\rm sub}=0.46_{-0.22}^{+0.38}\times 10^{14}\,h_{70}^{-1}M_\odot$, respectively.  
The resulting halo ellipticity and orientation angle for the subcluster are $\varepsilon=0.47_{-0.30}^{+0.30}$ and $\psi_{\varepsilon}=38_{-47}^{+43}\,{\rm deg}$, respectively, where the orientation angle of the major axis, $\psi_{\varepsilon}$, is measured from north to east. 

We present a comparison with the hydrostatic mass of $M_{500}^{\rm H.E.} = 2.39 \pm 0.39 \times 10^{14}\,h_{70}^{-1} M_\odot$ \citep{2023MNRAS.520.6001P}, determined using the conventional X-ray analysis method. Initially, we identified an X-ray centroid and then extracted the surface brightness and temperature profiles. These profiles were subsequently fitted with analytical models, without taking into account any particular model of the X-ray substructure. In contrast, 2D WL analysis simultaneously determined WL masses of the two components and their centers. For a mass comparison, there are two approaches to estimate $M_{500}$ from the resulting masses. First, We estimate the total mass as the summation of each $M_{500}$ estimated from $M_{200}$ and obtain $M_{500}=M_{500}^{\rm main}+M_{500}^{\rm sub}=2.21_{-0.84}^{+0.84} \times 10^{14}\,h_{70}^{-1}M_\odot$. In this context, we estimate errors using the posterior distributions, which allows for the consideration of error correlations. Second, we compute $M_{500}(<r_{500})$ within the radii, $r_{500}$, from the WL-determined main center, where the mass of the subcluster is estimated by considering an off-centering effect in the three-dimensional space \citep{2021MNRAS.501.1701O} with the assumption of no separation along the line-of-sight because the redshifts of the BCGs of the main cluster ($z=0.12823$) and subcluster ($z=0.12869$) are almost the same (Figure \ref{fig:map}). The resulting mass is $M_{500}(<r_{500})=M^{\rm main}(<r_{500})+M^{\rm sub,off}(<r_{500})=2.08_{-0.77}^{+1.00}\times 10^{14}\,h_{70}^{-1}M_\odot$. Either case is consistent with $M_{500}^{\rm H.E.}$ within $1\sigma$ errors, though there remains systematic uncertainty of central positions.

For a visual purpose, the best-fit projected mass density of the elliptical NFW model in unit of $10^{15}\,h_{70}M_\odot\,{\rm Mpc}^{-2}$ is represented by white contours in the top panel of Figure \ref{fig:map}. The WL-determined centers are $(\alpha_{\rm main}, \delta_{\rm main})=(29.310_{-0.010}^{+0.005},-5.865_{-0.005}^{+0.005})$ and $(\alpha_{\rm sub}, \delta_{\rm sub})=(29.399_{-0.017}^{+0.009},-5.796_{-0.012}^{+0.009})$ for the main and sub cluster, respectively. The center of the main cluster coincides with the X-ray centroids of the main cluster but is slightly offset from the bright galaxy. The center of the subcluster is offset from the X-ray peak and the bright galaxy. The projected distance between the main and sub cluster is $923\pm137\,h_{70}^{-1}{\rm kpc}$, which is about a half of the sum of the overdensity radii, $r_{200}\simeq1985\,h_{70}^{-1}{\rm kpc}$.

When we also consider a single NFW model associated with the main cluster, the Akaike's information criterion (AIC) and Bayesian information criterion (BIC) become worse by $+10^4$, and thus the two mass components are better suited to describe the two-dimensional shear pattern.

\begin{figure*}[Htb]
\includegraphics[width=\hsize]{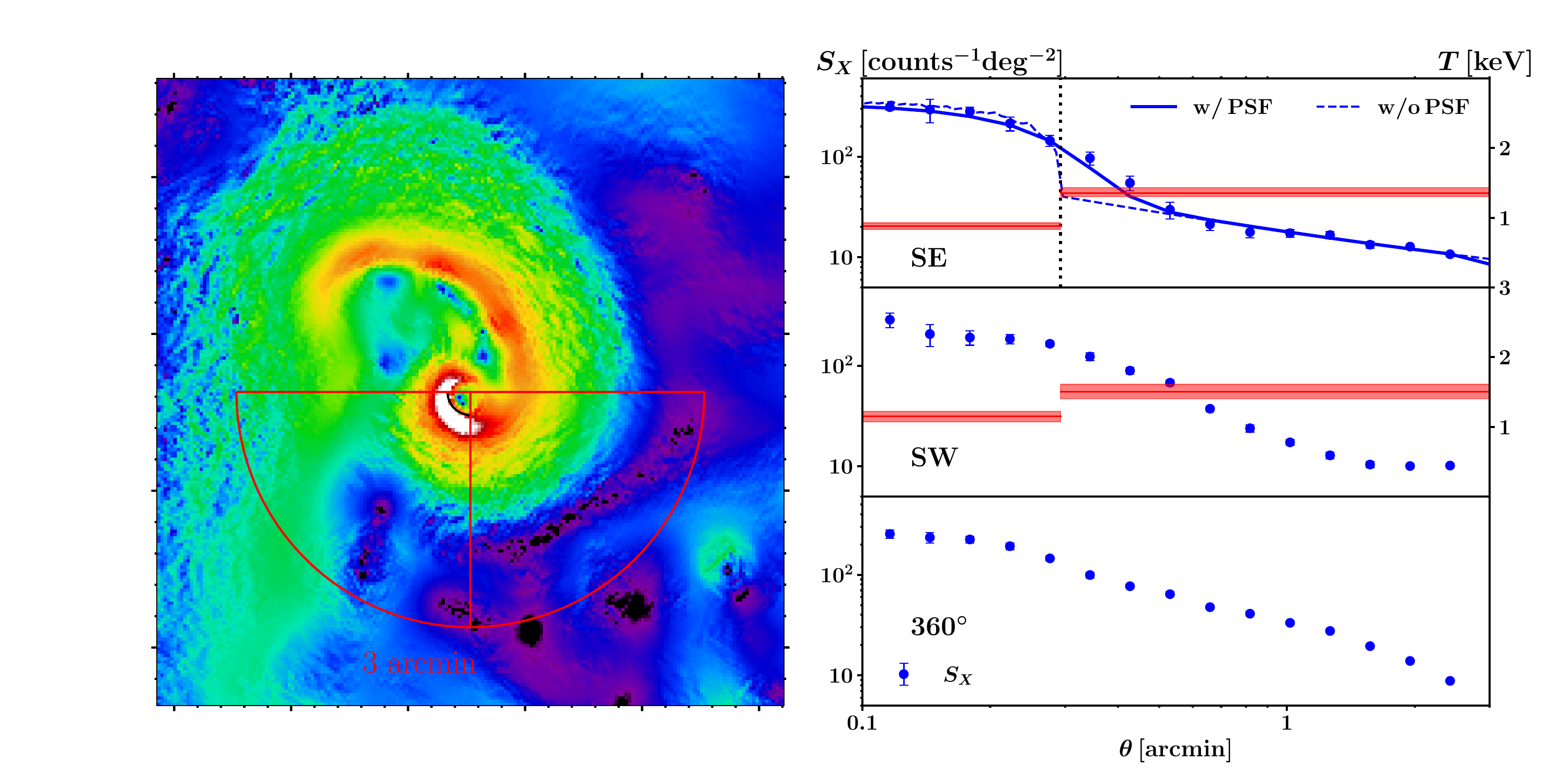}
\caption{{\it Left}: Adaptively GGM-filtered image ($8'\times 8'$). A comma-shaped tail is found in the northern-east (NE) direction. The red curve represents a maximum radius (3 arcmin) to compute surface brightness and X-ray temperature. The black curve denotes the best-fit discontinuity in the surface brightness profile. {\it Right}: Surface brightness and temperature profiles. From top to bottom, southern-east (SE), southern-west (SW), and the whole regions are represented. Blue points are the surface brightness profile in arbitrary unit. Blue solid and dashed lines represent the best-fit model convolved with and without the PSF, respectively. The red solid and regions are the temperature and its uncertainty, respectively. {Alt text: Two line graphs showing the X-ray count rate around the comma-shaped gas. }}
\label{fig:sxprof}
\end{figure*}

We examine the X-ray structure of the comma-like structure. The left panel of Figure \ref{fig:sxprof} shows adaptively GGM-filtered image around the comma-shape structure. We find a clear spiral pattern and an edge in the southern-east region of the dense core, indicating that the subcluster gas has large angular momentum. When we fit it with a general Archimedean spiral specified by $r=a+b \theta^{1/c}$, we obtain $a=32.7\,{\rm arcsec}$, $b=0.01\,{\rm arcsec}$, and $c=0.2$. Given the parameters, the angular rotation is weakly dependent on the radius, $\propto r^{0.2}$. The surface brightness profile in the southern-east direction is shown in the left-top panel of Figure \ref{fig:sxprof}. We fit it with a double power-law model of $n_{e,{\rm in}}\propto r^{-p_1}\,(r<r_{\rm dis})$ and $n_{e,{\rm out}}\propto r^{-p_2}\,(r>r_{\rm dis})$ with the PSF convolution. The MCMC approach is utilized for model fitting. Here, $r_{\rm dis}$ is the discontinuity radius. The resulting parameters are $p_1=0.073_{-0.047}^{+0.109}$, $p_2=0.888_{-0.104}^{+0.135}$, and  $r_{\rm dis}=0.293_{-0.008}^{+0.004}\,{\rm arcmin}$, respectively. Interestingly, the best-fit discontinuity radius agrees with the GGM edge (the left panel of Figure \ref{fig:sxprof}). The ratio of the electron number densities of the two components, $n_{e,{\rm out}}/n_{e,{\rm in}}=0.25\pm0.06$, at the discontinuity radius. 
We measure X-ray temperatures around the discontinuities. Our segmentation includes four distinct regions: those located within and outside of the discontinuities, as well as the western region, which shares an identical radius (Figure \ref{fig:sxprof}). We conduct a simultaneous fitting including the spectrum obtained from the background-dominated region spanning 9 to 17 arcminutes with the center at coordinates ($\alpha, \delta)=(29.328, -5.852$). Total fit statistic for pgstat is 4260 with 4167 degrees of freedom and $\chi^2=4307$, leading to null hypothesis probability of $6\%$. The resulting temperatures are shown in Table \ref{tab:xspec}. The ratio of the projected temperature in the outer and inner components across the discontinuity is $T_{\rm out}/T_{\rm in}=1.54\pm0.11$. By considering redshift as an unconstrained variable, applicable across the four regions, we derive a value of $z=0.096_{-0.039}^{+0.019}$, which is consistent with the galaxy redshifts (Figure \ref{fig:map}) at $1.7\sigma$ level.

\begin{table}[]
    \centering
    \begin{tabular}{c|cc}
         &  $k_B T$ &  Z\\
         \hline
SE inner &  $0.88_{-0.05}^{+0.05}$ & $0.31_{-0.14}^{+0.80}$\\
SE outer   & $1.35_{-0.05}^{+0.08}$ & $0.16_{-0.03}^{+0.04}$\\
 SW inner   &    $1.15_{-0.08}^{+0.07}$ & $0.24_{-0.07}^{+0.15}$\\
 SW outer   & $1.50_{-0.10}^{+0.11}$& $0.18_{-0.03}^{+0.04}$ \\
    \end{tabular}
    \caption{X-ray temperatures ($k_BT$) in units of keV and the metallicity ($Z$) in Apec model.}
    \label{tab:xspec}
\end{table}

\begin{table}[]
    \centering
    \begin{tabular}{c|ccc}
         &  no prior & prior & prior + \\
         &  &   &  X-ray morphology \\
      \hline   
     $\alpha_{\rm {main}}$    & $29.308_{-0.014}^{+0.006}$ & $29.310_{-0.010}^{+0.005}$ & $29.296_{-0.019}^{+0.011}$\\
     $\delta_{\rm {main}}$    &  $-5.865_{-0.006}^{+0.006}$ & $-5.865_{-0.005}^{+0.005}$  & $-5.864_{-0.005}^{+0.009}$\\
     $M_{200}^{\rm main}$    & $1.79_{-0.30}^{+1.61}$ & $2.68_{-0.93}^{+1.11}$  & $2.52_{-0.77}^{+1.11}$\\
     $\alpha_{\rm {sub}}$    & $29.402_{-0.012}^{+0.006}$ & $29.399_{-0.017}^{+0.009}$ & $29.404_{-0.012}^{+0.008}$\\
     $\delta_{\rm {sub}}$    & $-5.795_{-0.008}^{+0.008}$ &  $-5.796_{-0.012}^{+0.009}$  & $-5.799_{-0.018}^{+0.010}$\\
     $M_{200}^{\rm sub}$    & $1.17_{-0.43}^{+0.92}$ & $0.46_{-0.22}^{+0.38}$ & $0.41_{-0.18}^{+0.30}$\\
     $\varepsilon_\varepsilon^{\rm main}$    & - & $-0.01_{-0.30}^{+0.30}$  &  $0.02_{-0.31}^{+0.30}$ \\
     $\psi^{\rm main}$    & - & $2.83_{-0.94}^{+0.22}$ & $0.61_{-0.95}^{+0.97}$\\
     $\varepsilon^{\rm sub}$    &  -&$0.47_{-0.30}^{+0.30}$ & $0.47_{-0.29}^{+0.30}$ \\
     $\psi_\varepsilon^{\rm sub}$    & - &$0.66_{-0.82}^{+0.76}$ & $0.63_{-0.69}^{+0.88}$\\
       $y_{0,{\rm sub}}$    & - & -& $-0.37_{-0.22}^{+0.22}$\\
     $v_{0x}$    & - & -&$-921_{-261}^{+233}$\\
     $v_{0y}$    & -& -& $-257_{-152}^{+154}$
    \end{tabular}
    \caption{The resultant parameters of 2D WL analysis. The units of the parameters are as follows: $\alpha$ and $\delta$ are degrees, $M_{200}$ is $10^{14}\,h_{70}^{-1} M_\odot$, $\psi_\varepsilon$ is radian, $y_{0,{\rm sub}}$ is $h_{70}^{-1}$Mpc, and the initial relative velocity, $v_{0}$ is ${\rm km\,s}^{-1}$, respectively.  As for the projected ellipticity, negative values are allowed to avoid an artificial boundary at $\varepsilon=0$ and its absolute values are used for the constraints. The orientation angle, $\psi_\varepsilon$, is measured from north to east. Prior denotes the mass ratio prior from the stellar and BCG masses through baryonic scaling relation \citep{2022PASJ...74..175A}. X-ray morphology is an additional constraint from the X-ray filamentary structure and physical interpretation. The positive $x$ and $y$ directions to express the orbit point west and north, respectively.  
    We fix the initial position of the subcluster from the main cluster as $x_{0,{\rm sub}}=2\,h_{70}^{-1}\,{\rm Mpc}$.}
    \label{tab:mass}
\end{table}

\section{Discussion} \label{sec:dis}

We compare our result with numerical simulations from the Galaxy Cluster Merger Catalog website \citep{2018ApJS..234....4Z}. We use an off-axis merger simulation with a mass ratio of $1:10$ and an impact parameter of $b=1000$ kpc {and a spin parameter of 0.006.
\citep{2011ApJ...728...54Z}. The main cluster mass is $M_{200}=6\times 10^{14}M_\odot$, which is about three times higher than our WL mass. 
Although there are some uncertainties to compare with observational data, such as the presence of a cool core, the cool-core radius, the initial velocity and its direction, and the non-spherical structure, numerical simulations employing simplified physical conditions help aid in interpreting our findings. We take a snapshot where the subcluster is located at $\sim0.54r_{200}$ corresponding to the current position. The time is $1.8$ Gyr from the beginning and $0.6$ Gyr after the core passage. The X-ray surface brightness map is shown in the middle panel of Figure \ref{fig:ZuHone}. The merger occurs on the $xy$ plane, which is viewed from the z-direction. The white solid curve represents a trajectory of the subcluster in the rest frame of the main cluster. The subcluster moves from the bottom-right corner, passes through at the closest distance of $400$ kpc, and goes away to the top-left direction.  We fit the projected dark matter distribution with an elliptical NFW model and show the shape of the dark matter distribution of the subcluster as elliptical circles at $1.0$, $1.2$, and $1.4$, and $1.8$ Gyr, respectively. Despite an initially spherically symmetric mass distribution, the mass distribution is distorted during the merger, and its major axis is oriented towards the main cluster. The dark matter distribution is similar to our best-fit result. The projected ellipticity varies with its position by at most $\sim0.2-0.3$. It indicates that the subcluster is tidally deformed and rotating. The current comma-shape structure associated with the subcluster is found in the inset of the middle panel. The snapshots of the X-ray images at $1.2, 1.3, 1.4$, and $1.6$ Gyr are shown in the bottom panels of Figure \ref{fig:ZuHone}. The ram pressure stripping with angular momentum is found at $1.2-1.3$ Gyr. At the same time, the gas was also being pulled away by tidal rotation. Since the tidal rotation stripping is perpendicular to the direction of the major axis, the filamentary gas is distributed parallel to the trajectory of the subcluster. It is represented by the red auxiliary line. Once the gas has been stripped off by the tidal rotation, it is free of the tidal rotation and moves according to the two-body interaction. Thus, the gas filament aligns with the red auxiliary line.
These two features provide us with two important information. First, the tail direction of the comma-shaped structure and the edge positions are completely different from the merger moving direction, but local rotation. It assures us that we cannot infer the current moving directions of subclusters from X-ray morphologies. Second, the orientation angle of the filament is reflected by the trajectory around the pericenter.

The X-ray surface brightness profile perpendicular to the filament direction is shown in the second-top panel of Figure \ref{fig:ZuHone}. The distribution is expressed by a Gaussian profile with $\sigma\sim 100$ kpc. 

The X-ray surface brightness profile in the direction of the edge, like the SE direction of Figure \ref{fig:sxprof}, is represented in the top panel of Figure \ref{fig:ZuHone}. Here, we set the simulation at the cluster redshift.
The discontinuities of the surface brightness and temperature , $S_{X,{\rm out}}/S_{X,{\rm in}}\simeq0.1$ and $T_{\rm out}/T_{\rm in}\simeq1.7$, correspond closely with our measurements.

The main difference between the numerical simulations and our result is that the X-ray emission from the simulated main cluster is centrally concentrated and displays a sloshing pattern, whereas the X-ray emission from the observed main cluster itself is elongated and lacks the sloshing pattern.
This might be due to the triaxial shape of the halo  at the initial condition or to the presence of a non-cool core.

We next discuss the merger orbit based on the following two assumptions: (1) the cluster merger occurs on the sky plane for simplicity and (2) the motion of each cluster follows the two-body problem with Chandrasekhar's dynamical friction \citep{1943ApJ....97..255C}, as specified by 
\begin{eqnarray}
    \ddot{\bm r}_a &=&-\frac{GM_b(<r) \bm{r}}{r^3}+\bm{F}_{{\rm dyn},ab},\\
    \bm{F}_{{\rm dyn},ab}&=& -4\pi G^2 M_a(<r) \rho_{b}(r)\ln \Lambda (r) A_{{\rm vel},ab}\frac{\bm{v_{ab}}}{v_{ab}^3} ,\nonumber \\
    A_{{\rm vel},ab}&=&{\rm erf}(X_{ab})-\frac{2X_{ab}}{\sqrt{\pi}}e^{-X_{ab}^2}, \quad X_{ab}=v_{ab}/(\sqrt{2} \sigma_{v,b}),\nonumber 
    \end{eqnarray}
    for
    \begin{eqnarray}
    \ln\Lambda&=& \begin{cases}
    \ln \left(1+\frac{r^2}{p_{90}^2}\right)^{1/2}    & \hspace{-6em} (r\ge p_{90}=\frac{GM_a}{v_{ab,{\rm ini}}^2},M_a < M_b) \\
    \ln \left(1+\frac{r^2}{p_{90}^2}\right)^{1/4}\left(1+\frac{r^2}{p^2}\right)^{1/4} & \\
    +  \ln\left(1+\frac{M_b(<r)}{M_a(<r)}\right)&  ({\rm else})\\
      \end{cases}
\end{eqnarray}
where the subscript $a,b$ denotes the component index of each cluster, $\bm{r}_a$ and $\bm{v}_{ab}$ are the position and the relative velocity of the $a$ component, $\rho$ is a background mass density, $\sigma_v$ is a velocity dispersion calculated from the mass $M_b(<r)$ by Jeans equation with an isotropic dispersion, respectively. A Coulomb logarithm, $\ln \Lambda$, is described by $\ln \left(1+r^2/p_{90}^2\right)^{1/2}$ where $p_{90}$ is the impact parameter for a 90 degree deflection \citep{2008gady.book.....B}. In the time of $r/p_{90}<1$ for the subcluster, the simulated orbits are better depicted by incorporating a Coulomb logarithm into both the initial term ($p_{90}$) and local deflection term ($p=GM_a(<r_a)/v^2_{ab}$), alongside a traditional term \citep[e.g.][]{2016MNRAS.463..858P} that reflects a mass-dependent interaction efficiency. 
Employing this formula, the precision of the analytical solution remains approximately $0.1-0.3$ Mpc for up to 3 Gyr within simulations featuring mass ratios of $1/1$, $1/3$, and $1/10$, along with impact parameters of $b=0, 0.5$, and $1$ Mpc (the black dashed line in the middle panel of Figure \ref{fig:ZuHone}, see also \citet{2025arXiv251016291O}). We here ignore the dynamic friction due to non-uniform distribution.

We then constrain the merger parameters based on the aforementioned knowledge from numerical simulations, under the assumption that the merger takes place within the sky plane.
As orbital constraints, we use three conditions: (1) the orientation of the filament is the same as the tangential direction of the orbit at the pericenter, ${\mathcal L}_{\rm angle} \propto -0.5 \left(\frac{ \psi_{\rm fil}- \psi_{\rm peri}}{\sigma_{\psi}} \right)^2$, where $\psi_{\rm peri}=\arctan\left(-\frac{v_{x,\rm sub}}{v_{y,\rm sub}}\right)\bigr|_{\rm peri}$ represents the angle tangent to the path at the pericenter, with $(x_{\rm sub},y_{\rm sub})$ delineating the subcluster's position within a coordinate system where the positive directions for $x$ and $y$ are set to west and north, respectively, (2) the log-likelihood incorporates one of the orbital positions nearest to the current position, ${\mathcal L}_{\rm pos} \propto -0.5 \left(\frac{{\rm min}(|{\bm x}_{\rm pos}-{\bm x}_{\rm sub}(t)|)}{\sigma_{\rm pos}}\right)^2$, and (3) the trajectory intersects the filament, stratifying with $y_{\rm sub}(t)=f(x_{\rm sub}(t))$, where $f(x)$ represents a linear equation describing the filament. To consider the uncertainty of the WL masses and centers, we adopt a joint likelihood with the 2D WL shear data; ${\mathcal L}_{\rm angle} +{\mathcal L}_{\rm pos}+{\mathcal L}_{\rm 2DWL}$. We choose $x_{0,\rm sub}=2\,h_{70}^{-1}$ Mpc as an initial position of the subcluster and parameterize $y_{0,\rm sub}$, the initial relative velocity ($v_{0x},v_{0y}$) and the WL parameters. The resulting orbit is shown in Figure \ref{fig:modelmap}.

The resultant parameters are shown in Table \ref{tab:mass}. The best-fit parameters indicate that the initial impact parameter is $0.9$ Mpc, with an initial relative velocity of $956\,{\rm {\rm km\,s^{-1}}}$, corresponding to a spin parameter of $\lambda \simeq 0.026$. The resultant spin parameter is four times larger than that of the reference simulation ($\lambda=0.006$; Figure \ref{fig:ZuHone}). The velocity of the subcluster at pericenter (0.36 Mpc from the main cluster) is $2050\,{\rm km\,s^{-1}}$. The current velocity of the subcluster in the rest frame of the main cluster is $(-442, 1045)\,\mathrm{km\,s^{-1}}$ (the north-northeast direction, see the orbit in Figure \ref{fig:modelmap}) with a time of $0.7$ Gyr having passed after the pericenter. The WL masses are $M_{200}^{\rm main}=2.52_{-0.77}^{+1.11}\times 10^{14}\,h_{70}^{-1}M_\odot$ and $M_{200}^{\rm sub}=0.41_{-0.18}^{+0.30}\times 10^{14}\,h_{70}^{-1}M_\odot$, respectively.  The measurement accuracies improve by only $\sim10-20\%$ compared to the WL data alone. 

The current moving direction is proportional to the direction in which the gas tail is trailing, which indicates that the comma-like shape structure is not formed by ram pressure. Assuming the major axis of the subcluster is oriented along the north at the pericenter, the rotational energy within the discontinuity is estimated to be $E_{\rm rot}=(1/2)(2/5)M(<r_{\rm dis})r_{\rm dis}^2(62{\rm deg}/0.7{\rm Gyr})^2\sim\mathcal{O}(10^{51})\,{\rm erg}$, which is significantly lower than the initial total energy of $\mathcal{O}(|-10^{62}|)\,{\rm erg}$.

\begin{figure}[ht]
\includegraphics[width=\hsize]{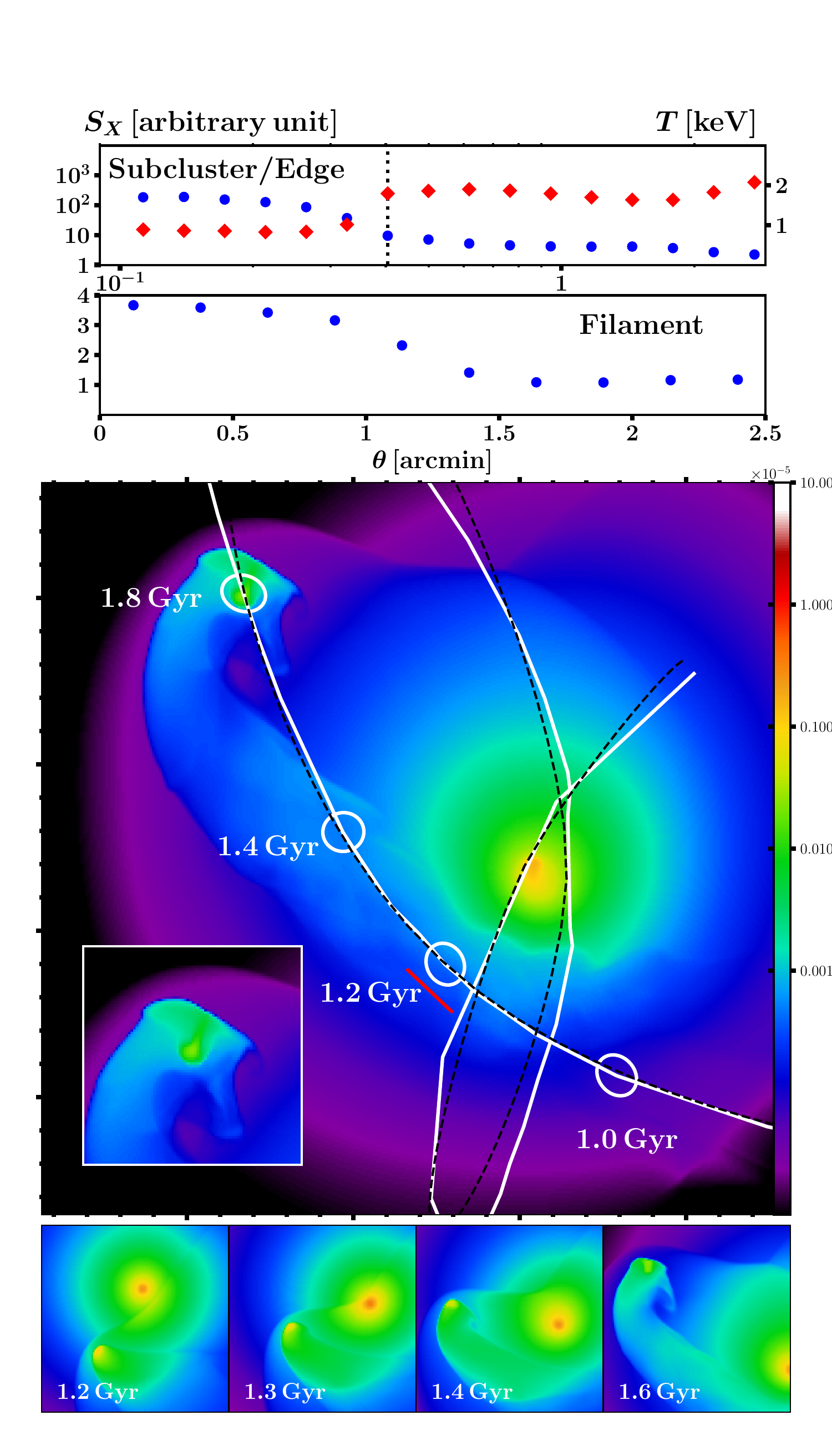}
\caption{Numerical simulation from \citet{2011ApJ...728...54Z} and \citet{2018ApJS..234....4Z}. {\it Middle}: the background colour represents the X-ray surface brightness at $\sim$1.8 Gyr after the beginning. The original image was inverted with respect to the x-axis direction. The spiral structure of the subcluster and the filamentary bridge between the main and sub clusters are found. The white solid and black dashed curves are a trajectory of the simulated subcluster and an analytical solution in the rest-frame of the main cluster, respectively. The white ellipticals represent the projected ellipticity of the DM halo at $1.0$, $1.2$, $1.4$, and $1.8$ Gyr after the beginning of the simulation (from right to left). The inset shows a zoomed view of the comma-shaped gas. The red line is an auxiliary line indicating the direction of the tidal rotation stripping. The red auxiliary line is parallel to the X-ray filament. {\it Top}: the surface brightness (blue circles) and temperature (red diamonds) profiles of the subcluster in the cold front (black dashed line) direction. {\it Second-top}: the X-ray surface brightness perpendicular to the filament. {\it Bottom}: The X-ray images at $1.2, 1.3, 1.4$ and $1.6$ Gyr. {Alt text: Four stacked graphs illustrating the results of the numerical simulation.}}
\label{fig:ZuHone}
\end{figure}

\begin{figure}
    \includegraphics[width=\linewidth]{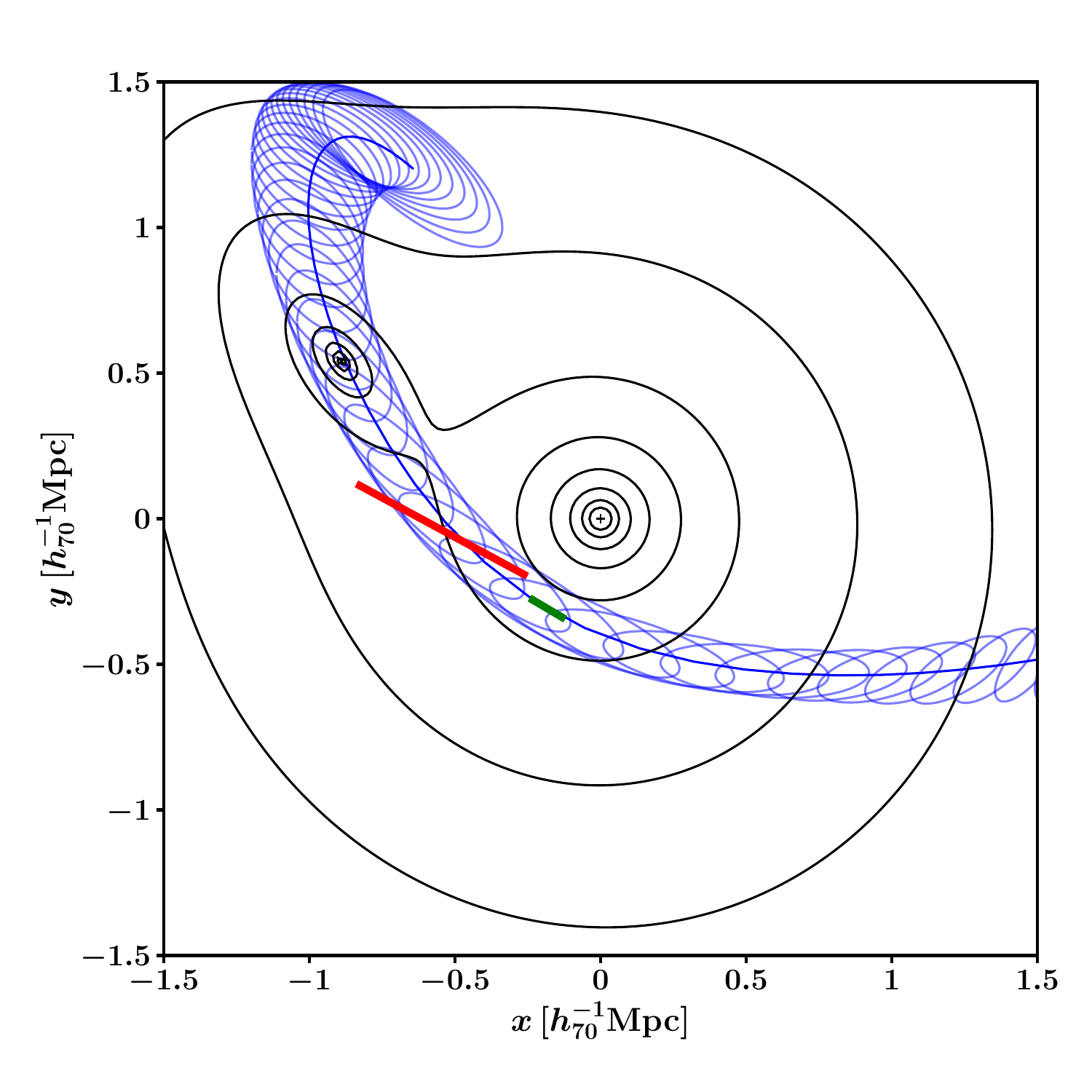}
    \caption{The model orbit. The blue curve and elliptical circles represent the best-fit orbit and the $1\sigma$ uncertainty at each time, respectively. The black contours are the projected mass distribution obtained by the joint constraint. The red solid line represents the X-ray filament. The green solid line is the tangential direction at the pericenter. {Alt text: A graph representing the results of the joint constraints. }}
    \label{fig:modelmap}
\end{figure}

\section{Summary} \label{sec:summary}

We found the X-ray filamentary structure between the comma-like subcluster and the main cluster with $\sim5.5$ significance level. 
The X-ray image of the comma-shaped subcluster, filtered adaptively by the GGM algorithm, follows the pattern of a typical Archimedean spiral.
Assuming prior information of stellar masses of the members and the BCGs, the 2D WL analysis yields $M_{200}^{\rm main}=2.68_{-0.92}^{+1.11}\times 10^{14}\,h_{70}^{-1}M_\odot$ and $M_{200}^{\rm sub}=0.46_{-0.22}^{+0.38}\times 10^{14}\,h_{70}^{-1}M_\odot$. The cluster is likely to be a minor merger. The main axis of the subcluster's projected mass distribution is more likely to align with the direction of the main cluster than to be oriented perpendicularly.

Similar X-ray and dark matter distributions are found in numerical simulations \citep{2011ApJ...728...54Z}. 
A qualitative scenario to form the filamentary structure is as follows. First, the gas stripped off by the ram pressure around the pericenter was pulled away by the tidal rotation. Second, the gas becomes free from tidal rotation and then inertially moves following the two-body interaction. Thereby, the X-ray filament is distributed parallel to the tangential velocity at the pericenter. In light of the situation, the X-ray filament, resulting from the cluster merger, in conjunction with the 2DWL results, enabled us to measure the parameters of the merger of the impact parameter and the initial infall velocity.

\section*{Acknowledgements}

This paper is based [in part] on data collected at the Subaru Telescope and retrieved from the HSC data archive system, which is operated by Subaru Telescope and Astronomy Data Center (ADC) at NAOJ. Data analysis was in part carried out with the cooperation of Center for Computational Astrophysics (CfCA), NAOJ. We are honored and grateful for the opportunity of observing the Universe from Maunakea, which has the cultural, historical and natural
significance in Hawaii.

This work made use of data from the Galaxy Cluster Merger Catalog (http://gcmc.hub.yt).

N.O. acknowledges JSPS KAKENHI Grant Number 25K07368.

\bibliographystyle{apj}
\bibliography{hscrefs}

\end{document}